\documentstyle[bbm,amsfonts,epsfig,12pt,here]{article}

\newcommand {\eqref} [1] {(\ref {#1})}
\newcommand {\slsh} [1] {\not{\hbox{\kern-2pt${#1}$}}}

\newcommand {\beq} {\begin{equation}}
\newcommand {\eeq} {\end{equation}}
  \newcommand {\ber}{\begin{eqnarray*}}
  \newcommand {\eer} {\end{eqnarray*}}
\newcommand {\bea}{\begin{eqnarray}}
  \newcommand {\eea} {\end{eqnarray}}

\newcommand{\Dslash}{\,{\raise.15ex\hbox{/}\mkern-12mu D}}

\begin{document}
\begin{titlepage}
\begin{flushright}{SWAT/07/505

UMN-TH-2528/06\,,  \,\,\, FTPI-MINN-06/38

CERN-PH-TH/2007-013\\}
\end{flushright}

\vskip 0.5cm

\centerline{{\Large \bf A note on \boldmath{$C$}-Parity Conservation and the}}

\vskip 0.2cm

\centerline{{\Large \bf Validity of Orientifold Planar Equivalence}}
\vskip 1cm
\centerline{\large Adi Armoni,${}^{a}$ Mikhail Shifman,${}^{b}$ and 
Gabriele Veneziano ${}^{c,d}$}

\vskip 0.3cm
\centerline{${}^a$ \it Department of Physics, Swansea University,}
\centerline{\it Singleton Park, Swansea, SA2 8PP, UK}
\vskip 0.2cm

\centerline{${}^b$   \it William I. Fine Theoretical Physics Institute,}
\centerline{\it University of Minnesota, Minneapolis, MN 55455, USA}

\vskip 0.2cm
\centerline{${}^c$ \it Theory Division, CERN}
\centerline{\it CH-1211 Geneva 23, Switzerland}
\vskip 0.1cm
\centerline{${}^d$ \it Coll\`ege de France, 11 place M. Berthelot, 75005 Paris, France}

\vskip 1cm

\begin{abstract}

We analyze the possibility of a spontaneous breaking of $C$-invari\-ance
in gauge theories with  fermions in vector-like ---
but otherwise generic --- representations
of the gauge group. QCD, supersymmetric Yang--Mills theory, and 
orientifold field theories, all belong to this class. 
We argue that charge conjugation is not spontaneously broken as long as
Lorentz invariance is maintained. Uniqueness of the
vacuum state in pure Yang--Mills theory (without fermions)
and convergence of the expansion in fermion loops
are key ingredients. The fact that $C$-invariance is conserved
 has an interesting application to our  proof of planar equivalence between supersymmetric Yang--Mills theory and orientifold field theory on $R^4$,  since it allows the  use of  charge conjugation to connect the large-$N$ limit of Wilson loops in different representations.

\end{abstract}

\end{titlepage}

\section{Introduction}
\label{introduction}

\noindent

There are very few tools that enable us to explore
QCD in the nonperturbative regime. Recently, building on an 
earlier idea due to Strassler \cite{Strassler:2001fs}, we suggested a new tool for analyzing nonperturbative QCD \cite{Armoni:2003gp,Armoni:2003fb}. We argued that one-flavor QCD can be approximated, within a $1/N$ error, by ${\cal N}=1$ 
super-Yang--Mills theory \cite{Armoni:2003fb} (for a review see \cite{Armoni:2004uu}). The relation between QCD and super-Yang--Mills was established by observing that 
SU$(N)$ gauge theory with the Dirac two-index antisymmetric fermion 
(to be referred to as the {\em orientifold field theory}) is nonperturbatively equivalent to super-Yang--Mills in a well defined bosonic subsector at $N\to\infty$. Planar equivalence  led to several strong predictions concerning QCD. Among them are the value of the quark condensate \cite{Armoni:2003yv} and the degeneracy of the $\sigma$ and the $\eta '$ mesons \cite{Armoni:2003fb} in one-flavor QCD. These predictions were supported by recent lattice simulations \cite{DeGrand:2006uy,Keith-Hynes:2006wm}.

In Ref.~\cite{Armoni:2004ub} we gave a formal proof of planar equivalence (see 
Ref.~\cite{Patella:2005vx} for a lattice strong-coupling version of the proof). 
Our proof assumes, implicitly,  charge conjugation invariance. The proof does {\em not} hold on compact spaces, such as $R^3\times S^1$,
as was first demonstrated in \cite{Barbon:2005zj}. 

Recently it was pointed out \cite{Unsal:2006pj} (for earlier works see
\cite{Kovtun:2004bz}) that a necessary and {\em sufficient} condition for orientifold planar equivalence to hold is 
the absence of  spontaneous breaking of  charge conjugation symmetry.
To further explore this observation the authors of \cite{Unsal:2006pj} considered the orientifold field theory on $R^3\times S^1$ (with a  small radius of $S^1$, so that the one-loop analysis can be trusted) and demonstrated that $C$-parity {\em is} spontaneously broken in this case, the order parameter being the Polyakov line in the
compactified direction.\footnote{In a revised version of the paper \cite{Unsal:2006pj}, it was argued that not only $C$-parity is broken, but $CPT$ as well.}
It was concluded that planar equivalence does not hold on $R^3\times S^1$, at least for
sufficiently small radii.\footnote{In fact, a more careful reading of 
\cite{Unsal:2006pj} implies that $C$ parity conservation and planar equivalence are invalid  only if periodic boundary conditions are imposed on the fermions. With antiperiodic boundary conditions,  both 
$C$ parity and planar equivalence  do hold, in contrast to claims otherwise 
\cite{Sannino:2005sk}. The sensitivity to the boundary conditions
is a clear-cut indication that this is a finite-size effect. The effects  found in \cite{Unsal:2006pj} are just Casimir-like effects that vanish as the theory decompactifies, i.e. as $R\rightarrow \infty$.} 

The above result was advertised (see e.g. the title of  \cite{Unsal:2006pj})
as raising doubts concerning the validity of planar equivalence on $R^4$. Shortly after,  the phase structure of the orientifold field theory on $S^3\times S^1$ was analyzed \cite{Hollowood:2006cq}. It was shown that, if the $S^3$ radius is sufficiently small so that perturbation theory can be trusted, the theory undergoes a phase transition as the 
$S^1$ radius increases. At large radius charge conjugation is restored. 
Although the analysis of \cite{Hollowood:2006cq} certainly cannot   be trusted in the domain of large $S^3$ radii, it could still  be considered as pointing in the opposite direction, i.e.  that planar equivalence holds on $R^4$. Another indication that planar equivalecnce holds on $R^4$ is provided by a recent lattice simulation \cite{degrand},  showing that QCD on a circle undergoes a phase transition from a $C$-parity violating phase to a $C$-parity preserving phase above a critical radius.

\vspace{1mm}

The purpose of this paper is to argue that $C$-parity does not break spontaneously in any vector-like gauge theory on $R^4$. Although we will not be able to give
a rigorous mathematical proof of the type known for 
spatial parity \cite{Vafa:1984xg}, we will present several convincing  physical arguments
that seem impossible to overcome.

In addition, we clarify certain aspects of our proof \cite{Armoni:2004ub} 
and point out which particular aspects of the set-up of Ref.~\cite{Unsal:2006pj}
are to blame for the failure of $C$-parity and planar equivalence 
on $R^3\times S^1$.

\section{\boldmath{$C$}-parity in pure Yang--Mills and \\
vector-like  gauge theories }
\label{C-parity}

In this section we will first argue that $C$-parity is not spontaneously broken in Yang--Mills theories
on $R^4$. We will then argue that, if $C$ is not broken in pure Yang--Mills theory,  it cannot be spontaneously broken if we add fermions in vector-like (real) representations.
 
The impossibility of  spontaneous breaking of $P$-parity at $\theta =0$
was proven long ago \cite{Vafa:1984xg}. This proof is in essence
nondynamical and is based only on certain general features
of Yang--Mills theories. Unlike spatial parity,
the issue of  spontaneous breaking of $C$ invariance
depends on the dynamics. This is the reason why we cannot
prove our assertion at the same level 
of rigor as that of Ref.  \cite{Vafa:1984xg}. Instead, we rely on a number of independent 
physical arguments which exploit known features of the gauge dynamics.

Consider first pure Yang--Mills theory on the cylinder $R^3 \times S^1$ with a large radius, and
the suspected order parameter for the spontaneous $C$-parity breaking, the Polyakov line in the compact direction (let us call it $t$),
\beq
P = {\rm Tr} \, \exp \left( i \int A_0 \, dt\right)\,.
\label{op}
\eeq
If $C$ is spontaneously broken, the vacuum expectation value (VEV)
of the Polyakov line will acquire an imaginary part, corresponding to two degenerate vacua with Im$\langle P\rangle = \pm K$ with $K$ a  non-vanishing constant. 

However, such a nonvanishing VEV would
contradict color confinement in pure Yang--Mills theory. 
Indeed, the order parameter (\ref{op}) is simultaneously an  order parameter for the  group's center. It must vanish in the confinement phase: and it does at low temperatures,
a well-established fact. There is a critical temperature below which color
confinement is recovered, and, correspondingly, $\langle P\rangle = 0$.

There is a subtle point in this argument. One can say that, as the radius of the
cylinder grows and eventually crosses the critical value beyond which
$\langle P\rangle = 0$, the Polyakov line no longer represents 
an appropriate order parameter for the spontaneous breaking of $C$ invariance of the theory.

Therefore, let us look at this problem from a more general perspective.
The Yang--Mills Lagrangian is $C$ invariant. Therefore,
for the spontaneous breaking to take place in pure Yang--Mills theory
its vacuum structure must be nontrivial. There should exist two
(more generally, an even number of) degenerate vacua
with  opposite $C$ parities.  

A certain amount of knowledge has been accumulated regarding
the vacuum structure of pure Yang--Mills theory.
From Witten's work \cite{Witten:1998uk} we know that the vacuum of pure Yang--Mills theory is nondegenerate at $\theta =0$. The vacuum energy is given by the partition function
\begin{eqnarray}
Z
&=&
\int DA_\mu \exp \left\{-
\int _{R^4}\,\, d^4 x \, \left [- {1\over 2g^2}\,  {\rm Tr}\, F_{\mu \nu} ^2 + i {\theta \over 16 \pi ^2}\, {\rm Tr}\, F\tilde F  \right] 
\right\}\nonumber\\[3mm]
&\equiv &\exp \left(-V\,E(\theta)\right) \, .
\end{eqnarray}
The vacuum at $\theta=0$ is the absolute minimum of energy with respect to other 
$\theta$ vacua since it is a sum of positive contributions (in the Euclidean formulation). 
What is of relevance for us is that it is also nondegenerate. The latter property can be seen explicitly at large $N$ by considering the string dual to pure Yang--Mills \cite{Witten:1998uk}.

The uniqueness (nondegeneracy) of the pure Yang--Mills vacuum
can be argued from a different side, starting from a slightly broken supersymmetric 
gluodynamics (i.e. ${\cal N}=1$ supersymmetric pure Yang--Mills theory).
This theory has $N$ degenerate vacua due to the spontaneous breaking of the 
$Z_{2N}$ chiral symmetry down to $Z_2$.
The order parameter is the gluino condensate
which was exactly calculated in \cite{mas}.

When a small mass $m/\Lambda \ll 1$ is given to the gluino
field,  the vacuum degeneracy is lifted  resulting in a theory with a nondegenerate 
vacuum. Pure Yang--Mills theory is recovered by taking the limit $m/ \Lambda \rightarrow \infty$. If the vacua of pure Yang--Mills theory were
degenerate, in order to end up with degenerate vacua  starting from 
a nondegenerate one, one would have 
a phase transition to occur at an intermediate value
of $m$. 
This is certainly impossible if the expansion in fermion loops (see below)
is convergent.

 Finally, 
 let us note that the uniqueness of the Yang--Mills vacuum is supported by lattice simulations.

Let us now include fermions.  In  vector-like theories
$C$ invariance is preserved at the Lagrangian level.
As in the previous case of pure Yang--Mills, spontaneous breaking
requires degenerate vacua. Taking for granted the uniqueness of the vacuum state
of pure Yang--Mills  (at $\theta =0$),  we may ask
what should happen in order that the inclusion of dynamical fermions introduces vacuum degeneracy.

Since fermions appear in the action bilinearly, it is easy to formally integrate them out.
The resulting theory is defined by the partition function
\beq
{\cal Z}=\int DA_\mu \, \exp \left\{
-S_{\rm YM} +\ln\, \det \left (  \slsh \!D + m \right )\right\} \, ,
\label{part} 
\eeq   
where $m$ is a mass parameter ($m\neq 0$, which can be viewed either as physical or  as an infrared regulator to be sent to zero at the end),
and the expression (\ref{part}) is given in  Euclidean space. 
Because of  the fermion determinant (which is taken in
the appropriate representation of SU($N$)$_{\rm color}$) the new effective action is nonlocal,
but this is irrelevant for our purposes. 

The expansion in fermion loops is just the expansion of the above (functional) integrand in powers of
 $\ln\, \det \left (  \slsh \!D + m \right )$ with the leading term
 corresponding  to the pure Yang--Mills theory. If such an expansion  
is convergent, the unique nondegenerate vacuum of pure Yang--Mills
is inherited by the theory with dynamical fermions. It is certainly the case  for 
sufficiently large $m$, as a result of the Appelquist-Carazzone theorem. A sudden switch to a divergent
regime, accompanied by occurrence of a vacuum degeneracy 
at a small value of $m$, is not supported by
any symmetry, and is highly unlikely whenever the inclusion of dynamical fermions does not spoil asymptotic freedom and confinement of the original Yang--Mills theory. One may still worry about the convergence
of the expansion\footnote{We thank M. \"{U}nsal and L. Yaffe for a discussion on this issue.} at $m=0$, since  one could suspect {\em a priori}
that infrared singularities would develop invalidating the
convergence.
However, also in this respect,  the two theories at hand, the one with and the one without fermions, behave similarly: indeed, there are no
massless particles in the physical spectrum of either
pure Yang--Mills or of the theory with a single fermionic  flavor.

Vacuum uniqueness implies no spontaneous
breaking of $C$. Note, that in the general case
the proof of planar equivalence is carried out at finite $m$
(then it becomes an equivalence between a non-supersymmetric theory and the 
one with softly broken supersymmetry \cite{Armoni:2004uu}).
Taking the limit $m\to 0$ at the end requires a special 
consideration. Infrared singularities due
to massless particles are potentially dangerous;
one must check that they do not contribute in an essential way.
To this end the limit $N\to \infty$ must be taken first.

\section{Implications for planar equivalence}
\label{PE}

\noindent

According to \cite{Unsal:2006pj}, the absence of spontaneous breaking of $C$-parity is a necessary and sufficient condition for the validity of orientifold planar equivalence. The necessity of this condition is absolutely obvious.
The very possibility of a spontaneous
breaking of $C$-parity is related to the possible existence
of two degenerate vacua of the theory  transforming into each other 
under the action of  $C$ parity. In pure Yang--Mills theory
the vacuum is unique.\footnote{For simplicity in this section we set $\theta =0$.}

Since Dirac quarks do not alter the conclusion --
charge conjugation cannot be broken on $R^4$  -- we can assert
that planar equivalence holds when the theory is formulated on $R^4$. 
In this section we wish to elucidate this issue
with regards to the ``refined" proof of Ref.~\cite{Armoni:2004ub}.

Let us recall  the main points of \cite{Armoni:2004ub}.  It was shown there that the partition functions of ${\cal N}=1$ super-Yang--Mills and the orientifold field theory, after integration over the fermions, coincide at large $N$. The corresponding derivation 
is based on the expansion of the partition functions
in powers of $\ln\det  \left ( i \slsh \!D - m \right )$ and then 
 expressing generic terms of the expansion  in terms of Wilson loops\footnote{It is essential to add a small quark mass, see comments at the end of this section.}.
Then we compare the results, term by term,
\beq
 \langle W_{{\rm SYM}} \, W_{{\rm SYM}}\, ...\, W_{{\rm SYM}} \rangle _{\rm conn.}= \langle W_{{\rm QCD}\mbox{-}{\rm OR}} \, W_{{\rm QCD}\mbox{-}{\rm OR}}\, ...\, W_{{\rm QCD}\mbox{-}{\rm OR}}\rangle _{\rm conn.} \, ,
 \label{equalp}
\eeq
at large $N$. The subscripts above are self-evident.

\vspace{1mm}

For example, for a single Wilson loop, Eq.~(\ref{equalp}) reduces to
\beq
 \langle W_{{\rm SYM}} \rangle = \langle W_{{\rm QCD}\mbox{-}{\rm OR}}\rangle \, . \label{equal}
\eeq 
The reason why Eq.~(\ref{equal}) holds is that, at large $N$,
\bea
&& \langle W_{{\rm SYM}} \rangle = \langle {\rm Tr}\, U\, {\rm Tr}\, U^\dagger \rangle = \langle {\rm Tr}\, U \rangle \langle {\rm Tr}\, U^\dagger \rangle \,,\\[3mm]
&& \langle W_{{\rm QCD}\mbox{-}{\rm OR}} \rangle = \langle {\rm Tr}\, U\, {\rm Tr}\, U \rangle = \langle {\rm Tr}\, U \rangle \langle {\rm Tr}\, U \rangle\,,
\eea
where $U$ is the Wilson loop in the fundamental representation. Under the assumption that $C$-parity is not broken, 
 $$  
 \langle {\rm Tr}\, U \rangle = \langle {\rm Tr}\, U^\dagger \rangle \,,
 $$ 
and  Eq.~(\ref{equal}) is obviously satisfied. The equality of all connected correlation functions  similarly follows \cite{Armoni:2004ub} under  the assumption that $C$-parity is not broken.
 
A closer look at our proof reveals that in fact we needed to make only the very mild assumption of no $C$-parity breaking in pure Yang--Mills theory (on~$R^4$).
This is due to the fact that 
all the above  Wilson loop correlation functions are calculated in the Yang--Mills vacuum, and not in the vacuum of the theory with quarks. How is the presence of quarks felt here? 

Since we expand in fermion loops, starting from pure Yang--Mills theory,
if this expansion is convergent, the number of vacua cannot change compared to that in pure Yang--Mills theory. The latter has a unique vacuum, and so does the theory with quarks defined by this expansion. 
In order to ensure the convergence of the fermion loop expansion on $R^4$ we  introduced the infrared  cut-off -- a mass term for the fermions. We believe that on $R^4$ it is sufficient, but we have to assume that the limit $m \rightarrow 0$ is 
nonsingular.
Practically, it is impossible to think otherwise, since there are
no massless particles in the theories under consideration at $m = 0$.
If the vacuum is unique,
certainly there is no place for the spontaneous breaking of $C$-parity.

There is another subtle issue which requires a mild assumption in our 
proof \cite{Armoni:2004ub}. We assume that the sum of the correlation functions commutes with factorization.  Lattice simulations can also shed light on  this point.

\section{Discussion and conclusions}

The result of Ref.~\cite{Unsal:2006pj}, and of the previous
investigation \cite{Cohen}, can be interpreted as follows.
The expansion in fermion loops does not exist in
the orientifold field theory on $R^3\times S^1$ with  periodic boundary conditions,
provided the $S^1$ radius is small enough. Inclusion of fermions
drastically changes the vacuum structure compared to that of pure
Yang--Mills theory. 
Technically, the spontaneous breaking of $C$-parity in the theory on $R^3\times S^1$,
demonstrated in \cite{Unsal:2006pj}, is associated with an order parameter that breaks $P$ invariance. Indeed, an expectation value of the Wilson line in the
compactified direction is generated. This is equivalent to 
the expectation value of a component of a vector.
Needless to say, this is an artifact of working with a Lorentz-non-invariant theory.
There are no lessons to be drawn from this analysis for Lorentz-invariant theories.

Our work was motivated by the potential importance
of the claimed nonperturbative  planar (large $N$)  equivalence between 
supersymmetric gluodynamics and the orientifold field theory (one-flavor QCD at $N=3$).
Such an equivalence, once firmly established, will provide a unique analytic tool in studying
both theories nonperturbatively. Although numerical methods are --- or will become
--- available for light dynamical fermions, the importance of making reliable analytic predictions can hardly be overestimated. 

\section*{Acknowledgments}

We are very grateful to Mithat \"{U}nsal and Larry Yaffe who pointed out a loophole in an argument presented
in the draft version of this paper.
A.A. thanks S. Elitzur, S. Hands, T. Hollowood, C. Hoyos, P. Kumar, A. Patella and B. Svetitsky for discussions. 
M.S. is grateful to T.~Cohen for useful exchange of opinions.
G.V. would like to thank L.~Giusti and E. Rabinovici for interesting discussions.

A.A.~is supported by the PPARC advanced fellowship award.
The work of M.S. is
supported in part by DOE grant DE-FG02-94ER408.

\end{document}